\documentclass[10pt,a4paper,twoside]{article}
\usepackage [latin1]{inputenc}
\usepackage{vmargin,fancyhdr,multicol,multirow,ifthen,cite,
            graphicx,wrapfig,calc,dcolumn,apalike,setspace,
            boxedminipage,rotating,textcomp,picinpar,longtable,
            url,amsmath,amssymb,ogonek} 

\usepackage[tight]{subfigure}
\usepackage{imc2022}

\begin{document}
\SetPaperBodyFont
\setlength{\footskip}{3.60004pt}

\begin{IMCpaper}{
\title{Tau Herculids 2022 from the McDonald Observatory, Texas - a progress report}
\author{N. O. Szab\'o$^1$, A. Igaz$^1$, L. L. Kiss$^{1,2}$, 
M. R\'ozsahegyi$^1$, K. S\'arneczky$^1$, B. Cs\'ak$^1$, L. Deme$^1$ and J. Vink\'o$^{1,2}$
        \thanks{$^1\,$Konkoly Observatory, CSFK Research Centre for Astronomy and Earth Sciences, \\
        Konkoly Thege \'ut 15-17, Budapest, 1121 Hungary \\
        $^2\,$ ELTE E\"otv\"os Lor\'and University, Institute of Physics, P\'azm\'any
    P\'eter s\'et\'any 1/A, Budapest, 1117, Hungary\\
                     \texttt{szabo.norton@csfk.org}}}

\abstract{As part of an intensive effort to observe the predicted 2022 Tau Herculids outburst, we recorded almost 800 individual meteor streaks on May 30th and 31st, 2022, using a high-sensitivity Sony~$\alpha$7 camera. The video recordings were obtained under perfect conditions at the McDonald Observatory, Texas, USA. The meteor sample is dominated by the predicted Tau Herculids shower, however, we also noted significant activity of sporadic meteors and other possible weak showers. We found that the time of the maximum activity matched very well the predictions, while we note the large fraction of faint meteors that were not detectable visually. We determined the radiant, and the time evolution of the activities and currently we are working on the determination of the brightness statistics.}}%


\vspace*{-3\baselineskip}
\section{Introduction}

Inspired by the predictions of a possible Tau Herculids outburst on May 31st, 2022 (see, e.g. Hori et al. 2008, Rao  2021), we have organised a dedicated field trip to the McDonald Observatory of the University of Texas at Austin, USA. The observatory is located in the western part of Texas, from where the expected radiant crossed the meridian just a few degrees from zenith, thus allowing nearly ideal observing conditions of the meteors. Moreover, the observatory is at the core of the Greater Big Bend International Dark Sky Reserve, where the most stringent lighting rules apply. We planned three consecutive nights of observations, centred on May 31st, 2022, so that in principle, we could have measured accurately the sporadic background on the preceding and the following night of the maximum. While this has not been achieved, the first two nights were perfect, which made our expedition fully successful.

Here we present a short progress report on the ongoing analysis of the whole dataset we gathered with three different types of video cameras. Of these, the very sensitive observations of a large number of faint members of the shower are the focus of this report.

\section{Observations}

The five observers on the field trip were A.I, M.R., L.L.K., L.D. and J.V., who watched the events visually and made parallel audio recordings on the night of the maximum. This helped us make a comparison between the visual impression and the video recordings. For the latter, A.I. used three different cameras as follows:

\begin{itemize}
    \item A DMK 33GX236 panel camera with a Computar 2.7-8 mm f/1.0 optics, for wide-field imaging of the sky over the southern horizon. This had the lowest sensitivity, roughly matching the visual limiting magnitude.
    \item A mobile AllSky7 camera system that has been designed to capture the full sky (Hankey 2021), with much better sensitivity than the small DMK camera.
    \item A Sony~$\alpha$7 camera with a Sony 24 mm f/1.4 optics for detecting the faint meteors in a limited field of view.
\end{itemize}

The Sony camera's sensitivity was set mostly to ISO 40,000, with which we reached a stellar limiting magnitude of about 8.5. The faintest meteor streaks were estimated to be about 0.5-1.0 magnitude brighter, indicating an excellent sensitivity in the visually invisible brightness range. The observations were recorded in 30 minutes long MP4 video files, with a frame rate of 25 fps (40 msec per frame), each file amounting to 11.25 GB of data. 

In total, we made 2 hours of observations on May 30th, i.e. the night before the maximum, so that we were able to test all the equipment and the local conditions next to the Visitor Center of the McDonald Observatory, where we installed the cameras. Then we continued with the whole night on May 31st, with 7 full Sony recordings that correspond to 3.5 hours of effective observing time with that camera. This has not been collected continuously, some gaps interrupted the observations, which ultimately spanned 5 hours in total on that night (from 3h UTC to 8h UTC). 

These two nights of Sony recordings form the basis of the current report. The analysis of the DMK and AllSky7 observations will be reported in a future study. 

\section{Data reduction}

We started with processing the MP4 video files of the Sony~$\alpha$7 camera. The first visual inspections of the recordings indicated that the videos contained a large number of faint and very short meteors around magnitude 8, appearing only on two of three individual frames. This has essentially disabled the application of conventional softwares for meteor analysis, given that most widely used methods would have confused these faint and short meteors with the noise in the video. Therefore, our approach was to use the superior pattern recognition capability of the human brain.  This has also been suggested by Peter C. Slansky (private communication), who has extensive hands-on experience with very similar cameras and meteor shower maxima (Slansky 2021). 

\begin{figure}[htb]
\centering
\includegraphics[width = \columnwidth]{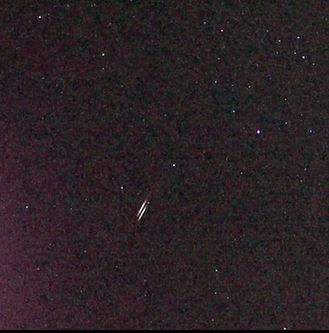}%
\vspace*{3pt}%
\caption{A twin meteor streak which illustrates the combined advantage of higher angular and temporal resolution with great sensitivity
offered by the Sony~$\alpha$7 camera. This particular example was recorded near the predicted maximum on May 31st. The field of view is approximately $5\times5$ degrees, the estimated brightness is about magnitude 3 for both meteors.}
\label{dupla-meteor}
\end{figure}

Our method to identify all meteors in the videos was to completely darken a room and watch the footage using QuickTime on an LG 55LA8609 flatscreen TV. The size of the video is too big for one person to fully capture all the sudden variations, so we split the screen into four quadrants and assigned one person to each quadrant. We also performed experiments on the optimal number of people needed to get the most meteors counted per video. We found that going from 3 to 4 people, there was a 50\% jump in the number of meteors noticed. However, a further increase in the number of watchers from 4-5 did not result in any further rise in the meteor counts. 

For the eleven 30 minutes long videos recorded over the timespan of two nights, we counted 796 meteors. Of these, we classified 626 as Tau Herculids and 170 as sporadic meteors. For every single meteor, we recorded the time of the appearance in the video to convert into UTC later, and the approximate position within the full frame (distance from the centre and position angle in hours from the top of the field). These steps were crucial to making later identifications easier.

With the time stamps of the meteors, we used a self-developed simple automatic command line interface of the VLC media player to select the individual frames containing the streaks. Subsequently, the frames were stacked by Siril using its Python integration. We used the offline database of {\tt Astrometry.net}\footnote{\tt http://astrometry.net} for converting the stacked frames into FITS files with WCS plate solutions so that the celestial coordinates of the trails became available. Finally, we manually went through the FITS images and measured by hand the RA and DEC coordinate pairs for the endpoints and the middle point of the meteor trail. After going through these steps, we ended up with 775 meteors with measured coordinates, implying that we achieved a 97\% success rate. Detailed inspection of several "lost meteors" revealed that they were caused by false positive spottings, failed plate solutions or unsuccessful identification.

\section{Radiant determination}

The radiant of the Tau Herculids was determined using the standard algorithm described by e.g. Schmitt (2004). The intersection point of each pair of the measured meteor trails was calculated in the following way. First, the RA ($\alpha$) and DEC ($\delta$) coordinates of the beginning and the end of each meteor trail was converted to Cartesian coordinates via
\begin{eqnarray}
x &=& \sin(\pi/2 - \delta) \cdot \cos(\alpha) \nonumber \\
y &=& \sin(\pi/2 - \delta) \cdot \sin(\alpha) \\
z &=& \cos(\pi/2 - \delta) \nonumber
\end{eqnarray}
Thus, ${\bf r}= (x, y, z)$ is the vector between a point on the meteor trail and the centre of Earth. 

The normal vector perpendicular to the plane defined by {\bf r}$_{\rm beg}$ and {\bf r}$_{\rm end}$ was computed as a vector cross product: ${\rm {\bf n}} = {\rm {\bf r}}_{\rm beg} \times {\rm {\bf r}}_{\rm end}$. The vector product of the normal vectors of each $i,j$ pair of the measured meteors defines the unit vector of the radiant:
\begin{equation}
{\rm {\bf R}} = {\rm {\bf n}}_i \times {\rm {\bf n}}_j.
\end{equation}

Finally, the Cartesian coordinates of the radiant, ($Rx$, $Ry$, $Rz$) were converted back to celestial coordinates as
\begin{eqnarray}
\alpha_R &=& \cos^{-1}(Rx / \sqrt{Rx^2 + Ry^2}) \nonumber \\
\delta_R &=& \sin^{-1}(Rz / \sqrt{Rx^2 + Ry^2 + Rz^2}).
\end{eqnarray}

This way we got a sample of radiants (actually, intersections of trails) for each pair of meteors in the sample. 

\begin{table}[]
    \centering
    \caption{The inferred coordinates of the Tau Herculid 2022 radiant. N gives the number of surviving meteors after 2 steps of sigma clipping. }
    \bigskip
    \begin{tabular}{lccccc}
    \hline
    Date & N & RA & error & DEC & error \\ 
       &  & (deg) & (deg) & (deg) & (deg) \\
    \hline
    2022-05-30 & 27 & 214.2 & 6.9 & 23.5 & 4.3 \\
    2022-05-31 & 224 & 213.1 & 4.0 & 28.6 & 2.4 \\
    \hline
    \end{tabular}
    \label{tab:radiant}
\end{table}

Since the presence of sporadic meteors may disturb the inference of the mean radiant of the shower, first, we filtered the input data by pre-selecting only those meteors whose trail pointed roughly toward the expected radiant in Coma Berenices. Instead of using all possible meteor pairs, the radiants of this sample were computed after pairing only the subsequent meteors. This way the scattering caused by the sporadic meteors that were still present in the sample was reduced substantially. Next, we applied a sigma clipping filter on the calculated intersections by removing those points that deviated more than 1 standard deviation ($1 \sigma$) from the mean value. After 2 iteration steps, the algorithm converged to a mean radiant with a standard deviation of a few degrees. The procedure is illustrated in Figure~\ref{fig:radiant} for both nights, and the coordinates of the inferred radiants for each night are shown in Table~\ref{tab:radiant}.

\begin{figure}[htbp]
\centering
\includegraphics[width=8cm]{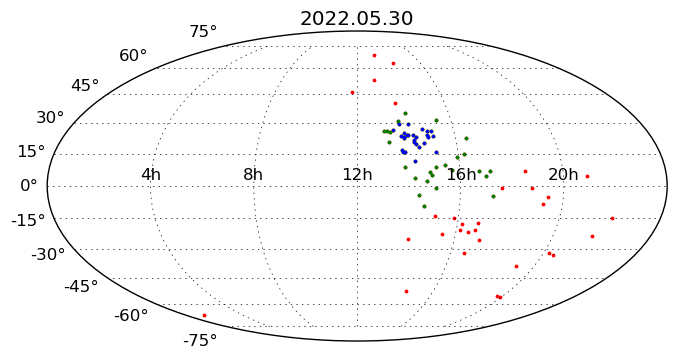}
\includegraphics[width=8cm]{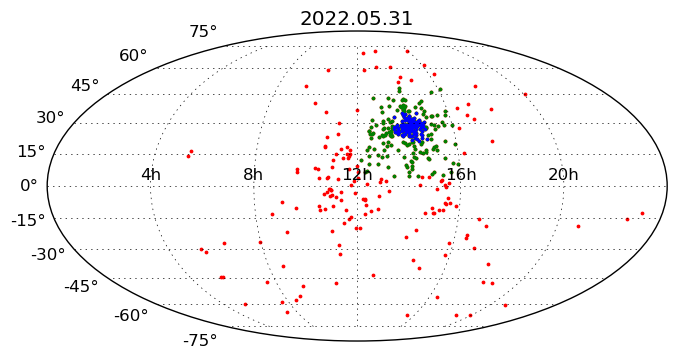}
\bigskip
\caption{The inferred intersection points of subsequent meteor pairs are plotted as red dots. Green dots mark the surviving points after the first sigma clipping iteration, while the blue dots denote the surviving points after the second iteration. The median and the standard deviation of the blue dots were assigned to the Tau Herculids apparent radiant and its uncertainty (see Table~\ref{tab:radiant}). }
\label{fig:radiant}
\end{figure}

\section{Conclusions}

Overall, the derived radiant coordinates agree very well with the predictions and the measured values published after the maximum. For example, looking at Table\ 1 of Ye \& Vaubaillon (2022), M. Maslov predicted the radiant to appear at (209.5$^\circ$, +28$^\circ$), while independent determinations from actual observations ranged from (209.17$^\circ$, +28.21$^\circ$) (Jenniskens 2022) to (208.6$^\circ$, +27.7$^\circ$) (Vida \& Segon 2022). Our values in Table\ 1 are identical to those results within the admittedly relatively large error bars. We find this agreement quite encouraging, given that our group has only recently started this line of observational investigations. 

The application of very sensitive and high-resolution video cameras, such as the Sony~$\alpha$7, for meteor observations, is a promising new development, however, the analysis of this kind of data is far from trivial. The huge data volume and the potentially large number of detections in the low S/N domain are beyond the capabilities of the currently used software tools and new directions, most likely using various Artificial Intelligence techniques, need to be explored.

Our project is still very much in progress. As of writing this report, we are exploring the possible ways of measuring meteor brightnesses, both from the individual and the stacked frames. The fine structure of the radiant, potentially depending on the meteor brightness is yet to be investigated. A combination of the full dataset from the three cameras may reveal changes in the population index between the brighter and the fainter ends of Tau Herculid distribution, which in turn would uncover important pieces of information on the properties of the dust cloud ejected from Comet 73P/Schwassmann-Wachmann~3 in the 1995 disintegration event. 

\section*{Acknowledgements}

The authors are grateful for the financial support from the E\"otv\"os Lor\'and Research Network (project ID K\"O-31 "Kozmikus hat\'asok \'es kock\'azatok"). We also acknowledge generous support and assistance from Judit Gy\"orgyey-Ries, N\'ora Egei and Zolt\'an Bels\H{o}.

\nocite{*}
\bibliographystyle{imo2}
\bibliography{ms}

\end{IMCpaper}
\end{document}